**Transport properties of c-oriented MgB$_2$ thin films grown by Pulsed Laser Deposition.**

C.Ferdeghini[a], V.Ferrando[a], G.Grassano[a], W.Ramadan[a], V.Braccini[a], M.Putti[a], P.Manfrinetti[b], A.Palenzona[b]

[a]INFM, Dipartimento di Fisica, Via Dodecaneso 33, 16146 Genova, Italy
[b]INFM, Dipartimento di Chimica e Chimica Industriale, Via Dodecaneso 31, 16146 Genova, Italy

**Abstract**
The electronic anisotropy in MgB$_2$, which arises from its layered crystal structure is not completely clear until now. High quality c-oriented films offer the opportunity of studying such property. MgB$_2$ thin films were deposited by using two methods both based on room temperature precursor deposition (by Pulsed Laser Ablation) and ex-situ annealing in Mg atmosphere. The two methods differ for the starting targets: stoichiometric MgB$_2$ in one case and Boron in the other. The two films presented in this paper are grown by means of the two techniques on MgO substrates and are both c-oriented; they present T$_C$ values of 31.5 and 37.4 K respectively. Upper critical field measurements, up to 9T, with the magnetic field in perpendicular and parallel directions in respect to the film surface evidenced anisotropy ratios of 1.8 and 1.4 respectively. In this paperwe will discuss this remarkable and surprising difference also in comparison with the literature data.

The discovery of superconductivity with T$_c$≈40K in a simple binary structure as the MgB$_2$ one [1] aroused tremendous interest for the fundamental and practical aspects of this material. MgB$_2$ exhibits a lot of intriguing properties: it is the compound with the highest T$_C$ among non-oxide superconductors and grain boundaries have not dramatic effects on the critical current densities [2, 3, 4], being the coherence length of this compound longer than those of HTSC and thus avoiding the depression of the superconducting order parameter between grains. These properties make MgB$_2$ a favorite candidate for large scale and electronic applications. The main limitation for large scale applications seems to be related to the considerable small value of the irreversibility field with respect to others technological superconductors as Nb-Ti [5].

Due to the layered structure of MgB$_2$ compound, an electronic anisotropy should be expected. A precise anisotropy determination requires quite large single crystals, not immediately available. An approximate evaluation was performed [6] by using aligned crystallites: the anisotropy factor η (i.e. the ratio between the critical fields parallel and perpendicular to the basal planes) resulted to be 1.7, as estimated from ac susceptibility measurements. In a hot pressed bulk sample η resulted to be 1.1 [7]. A surprisingly high η~6 value was estimated from conduction electron spin resonance and magnetization measurements performed on powder samples [8, 9]. Recently measurements on c-oriented thin films appeared in literature. Patnaik et al. [10] reported anisotropy measurements on films grown on SrTiO$_3$(111) starting from stoichiometric target: in three films with different residual resitivity ratio RRR ranging from 1 to 2, they found an anisotropy factor η in the range 1.8-2. M.H.Jung et al [11], instead, found a lower value η=1.25, despite they claim the epitaxiality of their film grown on $Al_2O_3$(1102) (starting from Boron precursor) with very low normal state resistivity and RRR=2.5. The very recent availability of small size single crystals does not definitively clarify this topic. In fact, the experiments performed on single crystals with RRR in the range 5-7, gave values in the interval 2.6-3 [12-14]. Therefore the actual η value needs to be confirmed being the reported values in a wide range.

The two films we present here were grown by Pulsed Laser Deposition. The PLD experimental apparatus consists of an UHV deposition chamber and a KrF excimer laser; details of the apparatus are described elsewhere [15]. We used for the first film (film 1 in the following) $MgB_2$ sintered target prepared by direct synthesis from the elements [16] and for the second one (film 2) a Boron target obtained by pressing amorphous B powders. Both precursor layers were deposited in high vacuum condition, at room temperature; film 1 was grown on MgO(100) and film 2 on MgO(111). We chose MgO substrates because they are very stable at the high temperature used in the subsequent annealing process, so minimizing the film-substrate interaction. The (100) crystallographic orientation has a square surface symmetry with a=4.203 Å.; the (111) orientation, instead, presents a hexagonal surface symmetry, with a lattice mismatch with $MgB_2$ less than 3%.

To crystallize the superconducting phase, we carried out an ex-situ annealing procedure in magnesium vapor. The samples were placed in a sealed tantalum tube with Mg lumps (approx 0.05 $mg/cm^3$), in Ar atmosphere, and then in an evacuated quartz tube and heated at T= 850°C for 30 minutes followed by a rapid quenching to room temperature.

From electrical measurements, we found $T_C$=31.4K with $\Delta T_C$=1.1K and RRR~1 for film 1 while film 2 exhibits $T_C$=37.5 K with $\Delta T_C$=0.6K and RRR=2.4.

To evaluate the structural properties of the samples, we performed standard x-rays θ-2θ analysis in the Bragg-Brentano geometry. The diffraction patterns show the presence of mainly (002) $MgB_2$ reflections with small peaks coming from different orientations and from other phases. The predominance of the (002) reflection, with respect to the other $MgB_2$ reflections, indicates that our films are preferentially c-oriented. Nevertheless the presence of the (110) $MgB_2$ reflection in both films indicates that a small fraction of this phase could be disordered. In any case, while in randomly oriented powders the tabulated intensity ratio between (002) and (110) reflections is about 0.5, we observe 0.8 for film 1 and 2.8 for film 2. Therefore we can conclude that the films are preferentially c-oriented and in particular film 2 is more oriented with respect to film 1. This fact is confirmed also from the rocking curves around the (002) peak reported in figure 1 for both films. It must be noted that the angular position of the (002) reflection for film 1 indicates larger c lattice parameter for this film. The curves have a full width at half maximum (FWHM) value of about 8° for the film 1 and about 3° for the film 2, confirming the preferential orientation of the samples.

Electrical resistance measurements as a function of temperature in applied magnetic field up to 9T were performed on both films in a Quantum Design PPMS apparatus by using a four-probe AC resistance technique at 7 Hz. The current density was always perpendicular to the magnetic field. $H_{c2}$ values were estimated from resistivity measurements. $H_{c2}$ vs. temperature curves were determined as the midpoint of the resistive transition for each field. In figure 2 we report $H_{c2}$ as a function of temperature for both films in the two orientations: $H_{c2}$ are considerably higher when the field is parallel to the film surface. In the same figure 2, as a comparison, the $H_{c2}$ of a sintered sample are also reported (the same bulk that we used as target for film 1).

We can observe, for film 1, a linear $H_{c2}$ versus T dependence near $T_C$, while for film 2 the upward curvature suggests a crossover between dirty and clean limit. The linear dependence found in the case of the film 1 can be accounted for the disorder and the impurities introduced during the growth.

However, by extrapolating at T=0, we can obtain for the two configurations:

For film 1: $H_{c2}^{par}$ = 26.4T and $H_{c2}^{perp}$ =14.6T.

For film 2: $H_{c2}^{par}$ = 19.6 T and $H_{c2}^{perp}$ = 13.8 T.

These values are considerable higher with respect to the bulk ones.

From these data it is possible to calculate the anisotropy factor η that resulted to be (at the lowest temperature) 1.8 for the film 1 and 1.4 for the film 2. These values are considerably lower in respect to the single crystal ones and this could derive from the not complete film orientation. Nevertheless we must consider that film 2 resulted to be more oriented with respect to film 1 but presents lower η; therefore the difference in η between the two films seems real and could be related with the preparation route. In fact the 1.8 value for film1 is in a good agreement with the results reported in [10] for films obtained from stochiometric target and the lower value for the film 2 agrees with the result reported in [11] for film obtained from Boron target, despite the referred samples are grown on different substrates. The preparation route can also determine the thin film chemical and microstructural characteristics: as an example, the boron precursor produces samples with higher RRR and higher $T_C$. Increasing the RRR factor (and $T_C$) we observe a crossover between dirty and clean limit condition, i.e. from a temperature independent to temperature dependent anisotropy factor.

In conclusion we studied the $MgB_2$ anisotropy factor in two different thin films finding a η dependence on the preparation route but a clear understanding of the dependence of the anisotropy factor from various parameters is still lacking.

**Figures caption**
Figure 1. Rocking curves around the (002) reflection for film1 and film2.
Figure2. $H_{c2}$ versus temperature for film1 (squares), film 2 (circles) for magnetic fields perpendicular (full symbols) and parallel (open symbols) to the film surface. $H_{c2}$ for polycrystalline bulk sample is also reported (asterisks).

**Fig1**

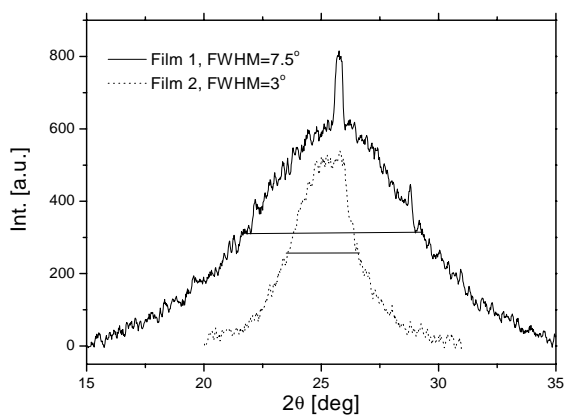

**Fig2**

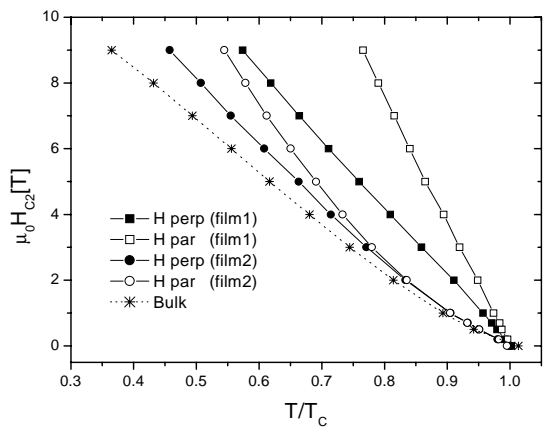